\documentstyle[aps,prb,multicol,epsf]{revtex}
\begin{document}

\title{The role of Zhang-Rice singlet-like excitations in one-dimensional cuprates}
       
\author{J.~Richter, C.~Waidacher, and K.~W.~Becker}

\address{Institut f\"ur Theoretische Physik,
  Technische Universit\"at Dresden, D-01062 Dresden, Germany}
  
\maketitle

\begin{abstract}
We present the first calculation of  the  electron-energy loss spectrum of 
infinite one-dimensional undoped CuO$_3$ chains within a multi-band Hubbard model.
The results show good agreement with experimental spectra of Sr$_2$CuO$_3$.
The  main feature in the spectra is found to be due to the 
formation of Zhang-Rice singlet-like excitations.
The ${\bf q}$-dependence of these excitations is 
a consequence of the inner structure of the Zhang-Rice singlet.
This  makes the  inclusion of the oxygen degrees of freedom  
essential for the description of the relevant excitations.
We observe that no enhanced intersite Coulomb repulsion is necessary 
to explain the experimental data.

\end{abstract}

\draft

\begin{multicols}{2}

Recently, charge excitations in the quasi one-dimensional compound
Sr$_{2}$CuO$_{3}$ have been investigated both experimentally$^{1-3}$ and
theoretically.$^{2-6}$ Sr$_{2}$CuO$_{3}$ is composed of
chains formed by CuO$_4$ plaquettes which share the corner
oxygens. The magnetic properties of these chains have been successfully
described using  a 
one-dimensional spin-$\frac{1}{2}$  Heisenberg antiferromagnet.$^{7-9}$ 

\begin{figure}
\epsfbox{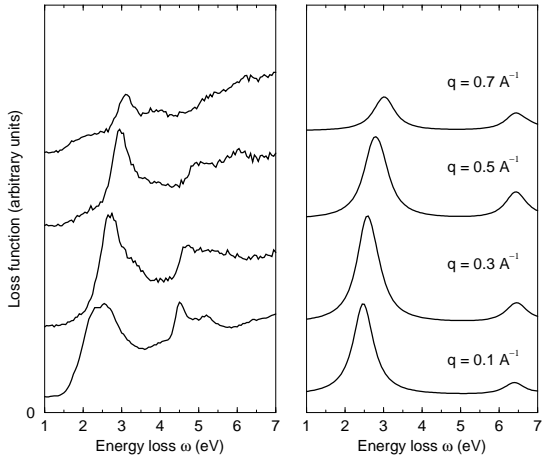}
\narrowtext
\caption{Comparison of experimental data for Sr$_2$CuO$_3$ (left), taken from 
Ref.~3, and the present theoretical results for the 
one-dimensional multi-band Hubbard model (right). The theoretical line spectra 
have been convoluted with a Gaussian function of width $0.1~\mbox{eV}$. 
For details see the text.
\protect\label{spectra}}
\end{figure}

Experimentally, the electron-energy loss spectrum (EELS) of Sr$_{2}$CuO$_{3}$
 \cite{neudert98} shows several interesting features (see Fig.~\ref{spectra}):
For small momentum transfer ($q=0.08$ \AA $^{-1}$) parallel to the
chain direction, one observes a broad peak around $2.4$ eV energy loss, and two
relatively sharp, smaller maxima at $4.5$ and $5.2$ eV. With increasing momentum
transfer, the lowest-energy feature shifts towards higher energy, reaching 
$3.2$ eV at the zone boundary ($q=0.8$ \AA $^{-1}$). Thereby its spectral width
decreases. In addition, the peaks at $4.5$ and $5.2$ eV lose spectral
weight as the momentum transfer increases, while some less well-defined
structures emerge around $6$ eV.

So far, these results have been compared only to calculations in an extended 
one-band Hubbard model.$^{3,6}$ From this comparison, 
Neudert {\it et~al.}$^{3}$ concluded that in Sr$_{2}$CuO$_{3}$ there is an 
unusually strong intersite Coulomb repulsion $V$: In the one-band model it is 
 of the order of 1 eV. 
 It is argued that  this large value of $V$ allows for the formation of 
excitonic states which are observed in the experiment. One of the aims of 
this paper is to show that no intersite Coulomb repulsion is necessary to 
explain the basic features of the experiment, if the O degrees of freedom 
are taken into account within the framework of a multi-band Hubbard model.

We investigate the EELS spectrum of a one-dimensional CuO$_{3}$ chain
system, using a multi-band Hubbard Hamiltonian at half-filling. In the hole 
picture this Hamiltonian reads

\begin{eqnarray}
H &=&\Delta\sum_{j\sigma}n^p_{j\sigma}
+U_{d} \sum_{i}n^d_{i\uparrow}n^d_{i\downarrow}\nonumber\\
&&+t_{pd} \sum_{<ij>\sigma}\phi^{ij}_{pd}
(p^\dagger_{j\sigma}d_{i\sigma}+h.c.)\nonumber\\
&&+ t_{pp} \sum_{<jj^\prime>\sigma}\phi^{jj^\prime}_{pp}
p^\dagger_{j\sigma}p_{j^\prime\sigma}~\mbox{,}\label{1}
\end{eqnarray}
where $d^\dagger_{i\sigma}$ ($p^\dagger_{j\sigma}$) create a hole 
with spin $\sigma$ in the $i$-th Cu $3d$ orbital ($j$-th O $2p$ 
orbital), while $n^d_{i\sigma}$ ($n^p_{j\sigma}$) are the 
corresponding number operators. The first and second term 
in Eq.~(\ref{1}) represent the atomic part of the 
Hamiltonian, with the charge-transfer energy $\Delta$, and 
the on-site Coulomb repulsion $U_{d}$ between Cu $3d$ 
holes. The last two terms
in Eq.~(\ref{1}) are 
the hybridization of Cu $3d$ and O $2p$ orbitals (hopping strength 
$t_{pd}$) and of O $2p$ orbitals (hopping strength $t_{pp}$).  The 
factors $\phi^{ij}_{pd}$ and $\phi^{jj^\prime}_{pp}$ give the correct 
sign for the hopping processes. Finally, $\langle ij \rangle$ denotes the summation over 
nearest neighbor pairs. 

The loss function in EELS experiments is directly proportional to the dynamical
density-density correlation function $\chi_{\rho}(\omega,{\bf q})$,
\cite{schnatterly77} which depends on  the energy loss $\omega$ and 
momentum transfer ${\bf q}$. $\chi_{\rho}(\omega,{\bf q})$ is calculated from
\begin{equation}
\chi_{\rho}(\omega,{\bf q}) = \frac{1}{i} \int_0^{\infty} dt~e^{-i\omega t}
\langle\Psi|
[ \rho_{-{\bf q}}(0),\rho_{\bf q}(t)]|\Psi\rangle~\mbox{,}\label{2}
\end{equation}
with 

\begin{equation}
\rho_{{\bf q}} = \sum_{i\sigma} n_{i\sigma}^d e^{i{\bf q}{\bf r}_i}
 +\sum_{j\sigma} n_{j\sigma}^p e^{i{\bf q}{\bf r}_j}~\mbox{,} 
\end{equation}
where $|\Psi\rangle$ is the ground 
state of $H$, and $\rho_{{\bf q}}$ is the Fourier transformed hole density.
The ground state $|\Psi\rangle$ is approximated as follows:\cite{waid2} 
 We start from a 
N\'eel-ordered state $|\Psi_N\rangle$ with singly occupied Cu $3d$ 
orbitals (with alternating spin direction) and empty O $2p$ orbitals.  
Fluctuations are added 
to $|\Psi_N\rangle$ using an exponential form
\begin{equation}
\label{3}
|\Psi\rangle = \exp \left(\sum_{i\alpha}\lambda_{\alpha}F_{i,\alpha}\right)
|\Psi_N\rangle~\mbox{.}
\end{equation}
The fluctuation operators $F_{i,\alpha}$ describe 
various delocalization processes of a hole initially 
located in the Cu $3d$ orbital at site $i$, where a summation over equivalent
final sites takes place.\cite{waid2} 
The parameters $\lambda_{\alpha}$ in Eq.~(\ref{3}) describe the 
strength of the delocalization processes and are determined 
self-consistently by solving the system of equations 
$\langle\Psi|{\cal L} F^\dagger_{0,\alpha}|\Psi\rangle= 0$, where 
${\cal L}$ is the Liouville operator, defined as 
${\cal L}A= [H,A]$ for any operator $A$. 
These equations have to hold if $|\Psi\rangle$ is the ground state of $H$. 

Using Eqs.~(\ref{2}) and (\ref{3}), we calculate the EELS spectrum
by means of Mori-Zwanzig projection technique.\cite{Mori65} For 
a set of operators $D_\mu$, the so-called dynamical variables, the 
following matrix equation  approximately holds
\begin{eqnarray}
\label{4}
\sum_{\gamma}\left[ z\delta_{\mu\gamma} - \sum_{\eta}
\langle\Psi|D^\dagger_{\mu} {\cal L} D_{\eta}|\Psi\rangle
\left(\langle\Psi|D^\dagger_{\eta} D_{\gamma}|\Psi\rangle\right)^{-1}
\right]\times\nonumber\\
\quad \times~\langle\Psi|D^\dagger_{\gamma}
\frac{1}{z-{\cal L}} D_{\nu}|\Psi\rangle = 
\langle\Psi|D^\dagger_{\mu} D_{\nu}|\Psi\rangle
~\mbox{,}
\end{eqnarray}
where $z=\omega+i0$. In Eq.~(\ref{4}) the set of dynamical variables 
was assumed to be sufficiently large so that self-energy contributions 
can be neglected. The set $\{D_\mu\}$ contains the dynamical variable 
$D_{0}=\rho_{{\bf q}}$. Therefore, by solving Eq.~(\ref{4}), an approximation
for Eq.~(\ref{2}) can be obtained. Besides $D_{0}$, the set includes 
$D_{\alpha}= \rho_{{\bf q}} F_{0,\alpha}$ for all $\alpha$. 
The  $F_{0,\alpha}$ are the fluctuation operators used in the ground 
state Eq.~(\ref{3}), without the summation over equivalent final
sites. We use altogether $12$ 
dynamical variables and observe good convergence of the spectral function.

\epsfxsize=0.4\textwidth

In Fig.~\ref{spectra} the obtained results are compared to the experimental 
spectra from Ref.~\cite{neudert98}. The parameters in the Hamiltonian 
are chosen as follows: $U_{d}= 8.8~\mbox{eV}$ 
and $t_{pp}= 0.65~\mbox{eV}$ are kept constant at typical values.\cite{3} 
The values of $\Delta=4.3~\mbox{eV}$  and $t_{pd}=1.5~\mbox{eV}$
have been adjusted to obtain the correct position of the lowest energy feature
at $2.5~\mbox{eV}$ for $q=0.01$ \AA $^{-1}$, and at
$3.1~\mbox{eV}$ for  $q=0.7$ \AA $^{-1}$. Thus, we effectively use only two
free parameters.
It is found that the value of $\Delta$ dominates the excitation energy, which 
increases with increasing $\Delta$. The
dispersion of the peak depends mainly on $t_{pd}$ with increasing dispersion
for increasing hopping parameter. As compared to the standard
value $1.3~\mbox{eV}$,\cite{3} $t_{pd}=1.5~\mbox{eV}$ is slightly enhanced, 
in agreement with recent results of band structure calculations.\cite{rosner99}

\epsfxsize=0.4\textwidth
\begin{figure}
\begin{center}
\epsfbox{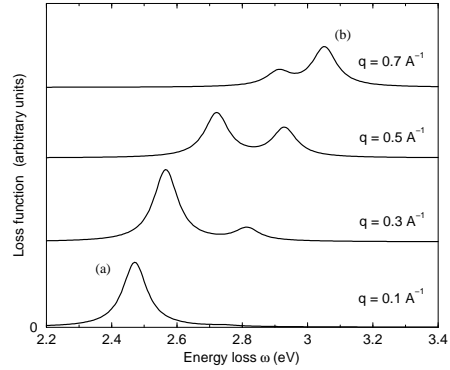}
\end{center}
\narrowtext
\caption{ The theoretical results for the dominant excitation at $2.4-3.1~\mbox{eV}$ 
with a broadening of $0.02~\mbox{eV}$. The  momentum dependence of the spectrum
is due to two different effects. First, with increasing ${\bf q}$ the spectral 
weight shifts from excitation (a) to (b). Second, the energies are 
${\bf q}$-dependent. 
Both effects contribute to about one half of the full momentum dependence. 
\protect\label{dispersion}}
\end{figure}

The theoretical spectra consist of two excitations. 
The dominant excitation is  at $2.45~\mbox{eV}$ for 
$q=0.1$ \AA $^{-1}$, and shifts to $3.05~\mbox{eV}$ for $q=0.7$ \AA $^{-1}$. 
Besides, a second  excitation appears at $6.4~\mbox{eV}$ which has no dispersion.

The low energy peak structure is shown in more detail in Fig.~2, where a smaller
peak broadening has been used. As will be explained below, mainly two different
Zhang-Rice singlet-like excitations \cite{zhang88} lead to this peak structure. 
The ${\bf q}$-dependence of the spectrum is due to two effects. Firstly, one observes 
a shift of spectral weight with increasing ${\bf q}$ between two excitations
labelled with (a) and (b) in Fig.~2. Secondly, with increasing ${\bf q}$  
the energies of the two peaks shift to higher values.

The shift of spectral weight
can be attributed to different delocalization
properties of the two final states.
The excited state (a) in Fig.~2 which dominates the spectrum for small 
momentum transfer is rather extended, see Fig.~3(a). This state has a  
rather small probability for the hole at its
original plaquette. With increasing {\bf q} the spectral weight 
shifts to another excited state, shown in Fig.~3(b), with
a higher  probability for the hole on its original Cu-site. 
This means that the character of the excitation 
changes from an extended to a more localized one, 
while still forming a Zhang-Rice singlet.

This behavior can be understood by analyzing the relevant 
expectation values in Eq.(5). For small values of {\bf q} the frequency term
$\langle\Psi|D^\dagger_{\mu}{\cal L} D_{\nu}|\Psi\rangle$ can be 
approximated  by expanding $e^{i{\bf qr}} \approx 1+ i{\bf qr}$ in
Eq.~(\ref{4}). 
This gives $\langle\Psi|F^\dagger_{0,\mu}{\cal L} F_{0,\nu}|\Psi\rangle\times
 {\bf q (r_{\mu}-r_{\nu})}$ which is proportional to the fluctuation distance, 
thus favoring far-reaching excitations. This picture changes for
large values of{ \bf q}, where stronger oscillations of the phase factor lead
to a cancelation of extended excitations. The result is a transfer of spectral weight 
from delocalized towards more localized excitations with increasing {\bf q}.

\epsfxsize=0.4\textwidth
\begin{figure}
\epsfbox{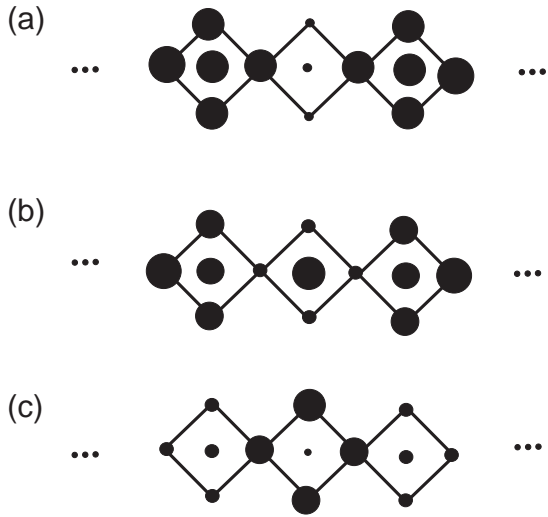}
\narrowtext
\caption{Hole delocalization properties of different excited states.
 Larger (smaller) dots symbolize a larger (smaller)
density of the hole originally located at the central plaquette.
States (a) and (b) are Zhang-Rice singlet-like excitations with different
delocalization properties. Excitation (a) has the largest spectral 
weight for small ${\bf q}$, whereas excitation (b) has dominant spectral 
weight for large ${\bf q}$, see also Fig.~2. Part (c) shows the local
excitation at $6.4$ eV, where the hole surrounds the central Cu site.
\protect\label{processes}}
\end{figure}

The ${\bf q}$-dependence of the energies, on the other hand, is a consequence 
of the inner structure of the Zhang-Rice singlet-like excitations.
In both excitations (a) and (b) a hole hops onto the Cu site of its 
nearest neighbor plaquette, see Fig.~3. Due to the Coulomb repulsion 
$U_{d}$, the hole which had originally occupied this Cu site is pushed 
away onto the surrounding O sites.
Depending on the direction of this delocalization, this process leads to a
${\bf q}$-dependence of the excitation energy. 

Next, we want to stress that the claim in \cite{neudert98} for the one-band
Hubbard model that only the inclusion of the next-neighbor repulsion leads 
to the possibility of the formation of an excitonic state is not consistent 
with our results. In the one-band model such a repulsion leads to a binding 
energy of empty and 
doubly occupied sites due to the reduction of neighboring interactions.
This binding energy is proportional to $V$. However, as can be seen from exact
diagonalization calculations in the one-band model,\cite{richter} the intersite 
repulsion mainly leads to an energy shift of the EELS spectra. Thus, the
parameter $V$ in the one-band model serves only to adjust the energetical 
position of the spectra, and is not necessary in more realistic models.
In the multi-band model, the formation of an exciton is only driven by the 
energetically favored formation of a Zhang-Rice singlet, and no further
inclusion of  next-neighbor repulsion is necessary.
 
The important role of the Zhang-Rice singlet formation has been studied 
previously also in an effective model for excitons in the CuO$_2$
plane.\cite{zhang98} Like the one-band model, this effective model neglects 
inner degrees of freedom of the Zhang-Rice singlet. If this model is 
reduced to the CuO$_3$ chain, ${\bf q}$-dependent energies are only possible 
for a non-vanishing O on-site Coulomb repulsion $U_p\neq 0$. In contrast 
to these results, we find ${\bf q}$-dependent energies for $U_p = 0$. As described 
above this effect cannot be explained in a model which neglects the 
inner structure of the singlet.

Thus, our results show that both an inclusion of the O-sites and a
complete description of the excitation is necessary to obtain the full 
dispersion. The O-sites are essential for the correct description of the
different characters of the singlet excitations, which leads to the shift of
spectral weight from one excitation to another.
On the other hand, taking account of the inner degrees of freedom of the
Zhang-Rice singlet leads to  the ${\bf q}$-dependence of the energies. 

The results of the projection technique do not correctly describe the
experimentally observed width of the peak for small
momentum transfer. A possible explanation is  that not all excitations are
included in the projection space. The above discussion suggests that
the width should be  due to the presence of additional delocalized
excitations. Processes which are neglected in the present calculation 
involve less important multiple excitations of holes beyond their
original plaquette.    

Finally, although they are not the focus of this paper, we discuss some
high-energy features. The excitation at 6.4 eV in the theoretical spectra is 
due to a local process on the plaquette itself. Here, the hole is excited to the 
surrounding O sites, without leaving its original plaquette, see Fig.~3(c). 
The energy of this structure does not shift as a function of momentum transfer. 
Once again, a transfer of spectral weight towards this localized excitation with
increasing values of {\bf q} is observed. The plaquette excitation  has a 
highly local character. Therefore, 
its spectral weight increases as a function of {\bf q} compared to the more 
delocalized Zhang-Rice singlet excitations. For small {\bf q} 
the spectral weight of the plaquette peak is about 6 times  smaller than that 
of the Zhang-Rice peak. As {\bf q } increases, this ratio increases to about 
one half. One should note that the experimental spectra show no obvious
features above 6 eV. However, since many different orbitals may contribute
in this energy range, we cannot expect a realistic description using a model 
that contains only Cu $3d$ and O $2p$ orbitals. This applies also to the experimental
structure around 4.5 eV for small momentum transfer, which is not described 
by the present model. We assume that this feature is due to excitations
which involve Sr orbitals, as has been argued before.\cite{neudert98}

In comparison with earlier works on Cu$2p_{3/2}$ X-ray photoemission
spectroscopy using the same theoretical approach,\cite{waid1} we find
that the character of the excitations in both experiments is very 
similar. Zhang-Rice singlet and local excitations play an
important role. In both experiments the dominant excitation at 
low energies is associated with a Zhang-Rice singlet formation. 

In conclusion, we have carried out the first calculation of the EELS-spectrum
for the one-dimensional CuO$_3$ chain by using a multi-band-Hubbard
model. Our results are in good agreement with experimental results for
Sr$_2$CuO$_3$. We find that the  main feature in the spectra is due to the 
formation of Zhang-Rice singlet-like excitations. The momentum dependence 
of the spectrum is due to two effects. First, there is a shift of 
spectral weight from less localized to more localized final states. 
Second, the excitation energies are ${\bf q}$-dependent. This ${\bf q}$-dependence is 
found to be a consequence of the inner structure of the Zhang-Rice singlet.
Therefore, the inclusion of the O degrees of freedom is 
essential for the description of the relevant excitations.
This has two important consequences. Firstly, only a multi-band model allows the
correct description of charge excitations. And, secondly, 
if a multi-band model is used, no intersite Coulomb repulsion is necessary. 
Furthermore, we observe the existence of a local excitation
at large ${\bf q}$-values.

We would like to acknowledge fruitful discussions with S.\ Atzkern, S.-L.\
Drechsler, J.\ Fink, M.\ S.\ Golden, R.\ E.\ Hetzel, A.\ H\"ubsch, R.\ Neudert, 
and H.\ Rosner. This work was performed within the SFB 463. 


\end{multicols}

\end{document}